\documentstyle[12pt,epsfig]{article}

\def\bd{{\rm B_d}}
\def\bs{{\rm B_s}}
\def\bc{{\rm B_c}}

\def\bpl{{\rm B^+}}
\def\bz{{\rm B^0}}
\def\bzbar{{\rm\overline {B^0}}}

\def\deltaGamma{{\Delta\Gamma}}
\def\deltaM{{\Delta\rm M}}
\def\deltaMd{{\Delta\rm M_d}}
\def\deltaMs{{\Delta\rm M_s}}
\def\ds{{\rm D_s}}

\def\dstar{{\rm D^*}}
\def\dstarpl{{\rm D^{*+}}}
\def\dstarmi{{\rm D^{*-}}}
\def\dz{{\rm D^0}}
\def\dzbar{{\rm\overline{D^0}}}
\def\dpl{{\rm D^+}}

\def\ks{{\rm K_s}}
\def\kl{{\rm K_L}}

\def\kpl{{\rm K^+}}
\def\kmi{{\rm K^-}}

\def\pipl{{\pi^+}}
\def\pimi{{\pi^-}}
\def\piz{{\pi^0}}
\def\psecinv{{\rm ps^{-1}}}
\def\rmix{{\rm r_{mix}}}
\def\rws{{\rm r_{ws}}}
\def\ra{{\rightarrow}}
\def\Vud{{\rm V_{ud}}}
\def\Vub{{\rm V_{ub}}}
\def\Vcd{{\rm V_{cd}}}
\def\Vcb{{\rm V_{cb}}}
\def\Vtd{{\rm V_{td}}}
\def\Vts{{\rm V_{ts}}}
\def\Vtb{{\rm V_{tb}}}

\def\Vubstar{{\rm V^*_{ub}}}

\def\Vcbstar{{\rm V^*_{cb}}}

\def\Vtbstar{{\rm V^*_{tb}}}

\def\Title#1{\begin{center} {\Large {\bf #1} } \end{center}}

\begin{document}

\title{Heavy Quark Lifetimes, Mixing and CP Violation} 
\author{\\ Guy~Blaylock\\ \\
{\it Department of Physics and Astronomy}\\
{\it University of Massachusetts}\\
{\it Amherst, MA 01003 U.S.A.}}
\maketitle
\thispagestyle{empty}
\setcounter{page}{0}
\begin{abstract}

This paper emphasizes four topics that represent
some of the year's highlights in heavy quark physics.
First of all, a review is given of charm lifetime
measurements and how they lead to better
understanding of the mechanisms of charm decay.
Secondly, the CLEO collaboration's new search for 
charm mixing is reported, which significantly 
extends the search for new physics in that sector. 
Thirdly, important updates in $\bs$ mixing are
summarized, which result in a new limit on $\deltaMs$,
and which further constrain the unitarity triangle.
Finally,  the first efforts to
measure CP violation in the B system are discussed.
Results are shown for the CDF and ALEPH measurements
of $\sin2\beta$, as well as the CLEO branching fraction
measurements of B$\ra$K$\pi,\pi\pi$, which have implications
for future measurements of $\alpha$.

\end{abstract}
\bigskip\bigskip\bigskip
\centerline{Presented at the}
\centerline{XIX International Symposium on Lepton and 
Photon Interactions}
\centerline{Stanford University, August 9-14, 1999}
\vfill\eject

\Title{\Large Heavy Quark Lifetimes, Mixing and CP Violation}

\bigskip\bigskip


\begin{raggedright}  
{\it Guy Blaylock\index{Blaylock, G.}\\
Department of Physics and Astronomy\\
University of Massachusetts,
Amherst, MA 01003}
\bigskip\bigskip
\end{raggedright}

\section{Introduction}

Much has happened this year in the subject of heavy quark studies.
This brief paper cannot hope to cover all of the details of work in 
this area, but will focus instead on a few of the highlights that have
emerged. First of all, recent high precision measurements in charm lifetimes,
particularly of the $\ds$, allow better understanding of the mechanism
of charm decays. Secondly, a new search for charm mixing at CLEO
significantly improves upon the sensitivity of previous analyses, and has
implications for the effects of new physics in the charm sector. Thirdly, significant
work has continued in the measurement of $\bs$ mixing, which puts important
constraints on the CKM parameter $\Vtd$. Finally, two of items of 
relevance to the study of CP violation in the B system have recently
been made available, which will be
touched on briefly.

\section{Lifetimes}

\subsection{charm lifetimes}

\begin{figure}[htb] 
\begin{center}
\epsfig{file=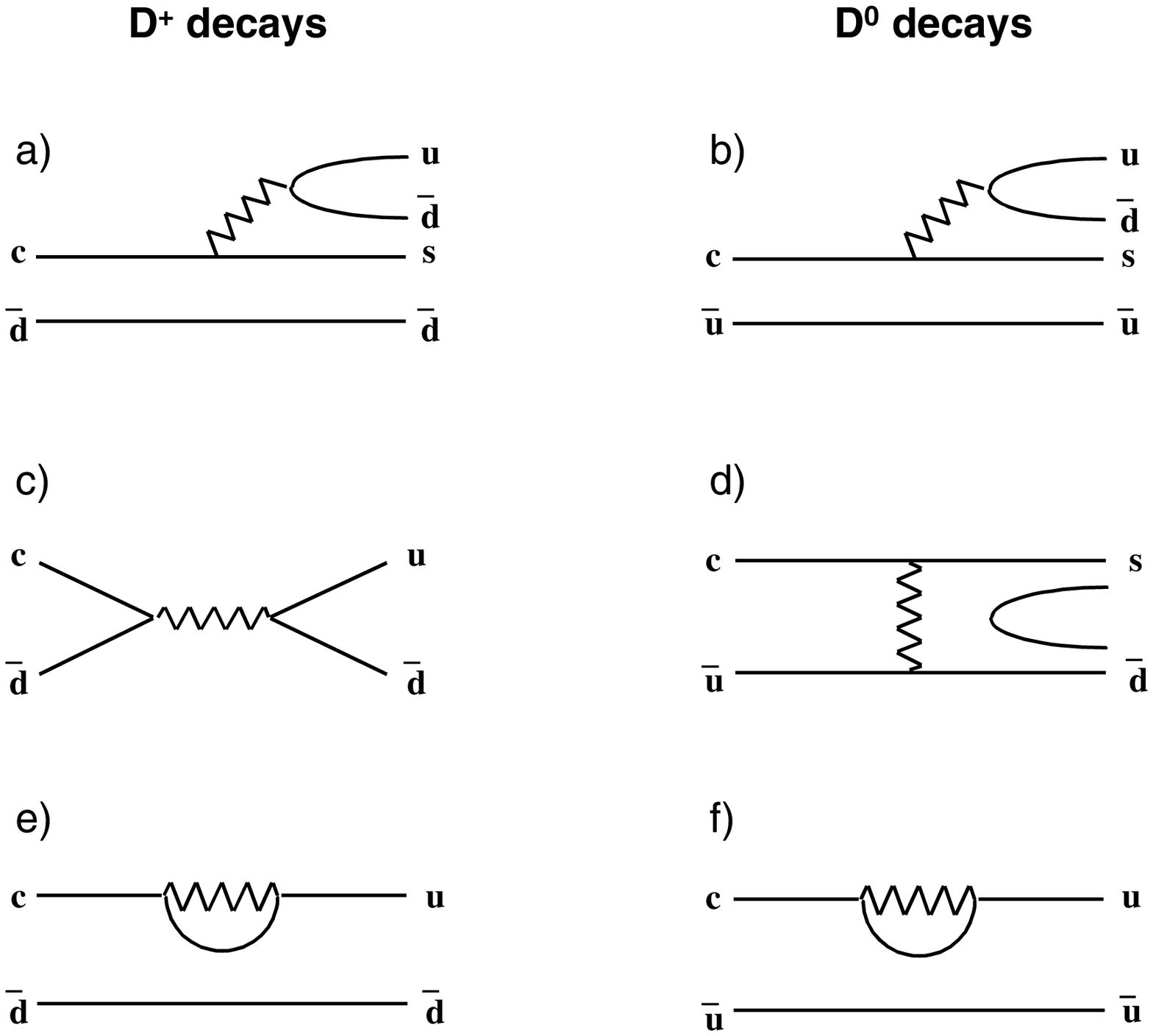,height=5.0in}
\caption{Diagrams contributing to $\dpl$ and $\dz$ decay.}
\label{fglifediags}
\end{center}
\end{figure}

To motivate the discussion of heavy quark lifetimes it is useful to recall an old puzzle in charm physics.
When researchers first measured the charged and neutral D meson lifetimes, they discovered 
that the $\dpl$ lifetime was considerably longer (about a factor of 2.5) than the $\dz$ lifetime. 
This result ran 
counter to expectations. 
The decays of both mesons were believed to be dominated by the spectator decay 
of the charm quark (Figures~\ref{fglifediags}a and \ref{fglifediags}b), which suggested
the two lifetimes should be nearly 
identical.


In the face of experimental evidence, several arguments were constructed as to why 
the two lifetimes might be different. First of all, the fact that there are two identical d quarks in 
the $\dpl$ final 
state (and not in the $\dz$) might give rise to Pauli-type interference, which could extend the 
lifetime of the 
$\dpl$. Moreover, other decay mechanisms, shown in Figures~\ref{fglifediags}c to \ref{fglifediags}f, 
were hypothesized, but these were 
expected to give only small contributions to the overall decay width. The $\dpl$ could decay via the weak 
annihilation of the charm and anti-down quarks (Figure~\ref{fglifediags}c) but this decay is Cabibbo suppressed, and therefore 
expected to be only a fraction of the spectator amplitude. Analogously, the $\dz$ could decay via the W 
exchange diagram of Figure~\ref{fglifediags}d, providing another difference between the two meson lifetimes. Both the 
weak annihilation and the W exchange amplitudes were expected to be small due to helicity and color 
suppression \cite{suppression}. However, both forms of suppression could be circumvented by the emission of a soft 
gluon, so the strength of the suppression was in question. Finally, penguin diagrams such as 
Figures~\ref{fglifediags}e 
and \ref{fglifediags}f could also contribute, 
but these diagrams are Cabibbo, helicity and color suppressed, and therefore 
received little attention.

In any case, the suggestion that mechanisms other than spectator quark decay could contribute significantly 
to the widths of the charm mesons provided motivation for further research in charm lifetimes. Primarily, 
this effort focussed on lifetime measurements of other weakly decaying charmed particles to use as 
comparison. Interference patterns were expected to be different for charm baryons, thereby providing a handle 
on the effects of Pauli type interference. This is especially easy to see in the case of the $\Omega_c$, where there are 
two identical strange quarks in the initial state. Moreover, W exchange is different for baryons, where the 
three quarks in the final state guarantee that the decay is neither helicity nor color suppressed. Finally, weak 
annihilation of charm and anti-strange quarks in $\ds$ decay is not Cabibbo suppressed, offering the 
possibility of studying this contribution.

\begin{figure}[htb] 
\begin{center}
\epsfig{file=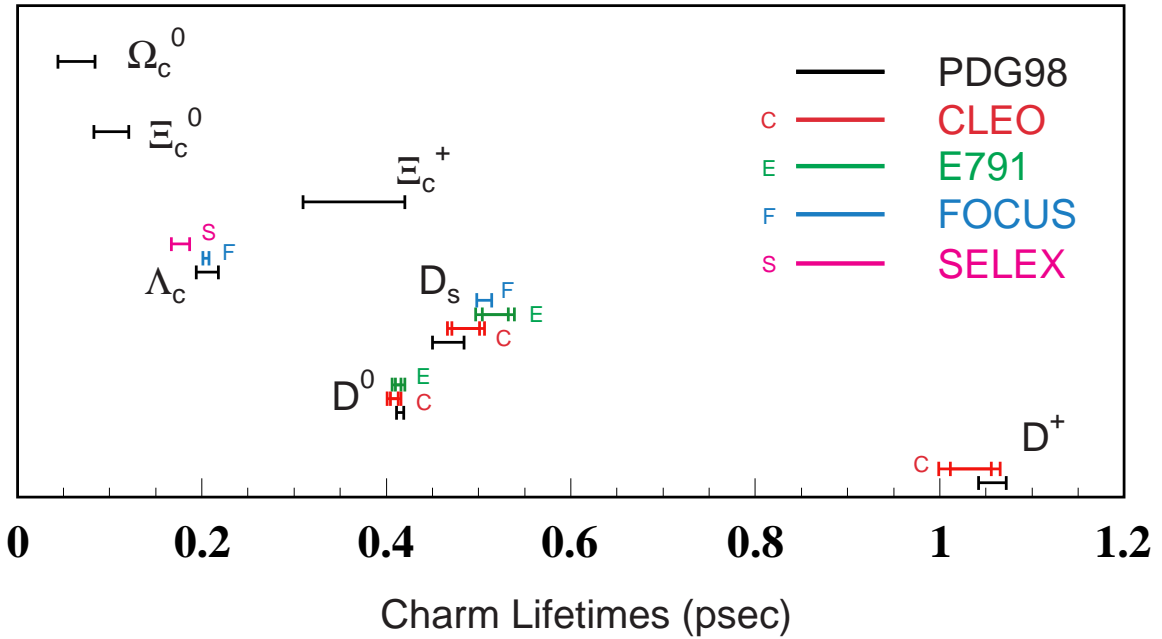,width=6.0in}
\caption[ ]{Measured lifetimes for all seven weakly decaying charmed particles.}
\label{fgcharmlife}
\end{center}
\end{figure}

Figure~\ref{fgcharmlife} shows where we stand today in the measurement of charm lifetimes. All seven weakly decaying 
charmed particles are shown. The unlabeled error bars give the world averages from the 1998 PDG review \cite{Caso:1998}. Since 
then, new measurements have been made available on $\dz$ and $\ds$ lifetimes from E791 \cite{Aitala:1998ca}, on all the D 
mesons from CLEO \cite{Bonvicini:1999ub}, and preliminary results on $\ds$ and $\Lambda_{\rm c}$ from FOCUS and SELEX 
\cite{Cheung:1999tz}. The most notable feature of the plot is that the $\ds$ and $\dz$ lifetimes are measurably
different. One year ago, these two lifetimes were nearly the same within errors. The new 
measurements not only reduce the error on the $\ds$ lifetime, but also shift the 
central value. 
An average of currently available data, including preliminary results 
from FOCUS yields \cite{Cheung:1999tz} $\tau_\ds/\tau_\dz = 1.211\pm0.017$. 


The precision of these new lifetime measurements, and the promise of more to come from the Fermilab 
fixed target experiments, finally allows us to study charm decays in a manner we have been wanting to do 
for 20 years \cite{lifetechnique}. As a simple example of how these data can be used to unravel the contributions to 
charm particle decay, consider the following three-step exercise to estimating Pauli interference, W exchange 
and weak annihilation contributions to D meson decays. 

First of all, compare the doubly-Cabibbo suppressed decay $\dpl\ra\kpl\pipl$$\pimi$ to its Cabibbo favored 
counterpart $\dpl\ra\kmi\pipl\pipl$. Since the kinematics are nearly the same in the two cases, the decays differ 
in only two respects. 
First of all, the decay diagrams have different (well-known) weak couplings.
Secondly, the Cabibbo-favored decay is subject to Pauli interference, while
the DCS decay is not, since there
are no 
identical particles in the DCS final state.
The ratio of the two rates can therefore be expressed:
\begin{equation}
{BR(\dpl\ra\kpl\pipl\pimi)\over BR(\dpl\ra\kmi\pipl\pipl)} 
\approx {\Gamma_{SP}\over\Gamma_{PI}}\times \tan^4\theta_C,
\label{eqSPtoPI}
\end{equation}
where $\Gamma_{PI}$ represents a spectator decay rate for the charm quark
that is subject to Pauli interference, while $\Gamma_{SP}$ represents a spectator decay rate without interference.
A combination of current measurements, including preliminary FOCUS data,
yields
\cite{Cheung:1999tz} ${BR(\dpl\ra\kpl\pipl\pimi)/BR(\dpl\ra\kmi\pipl\pipl)}=0.68\pm0.09\%$. Using this value, together with an estimate
for $\tan^4\theta_c$ of $2.56\times 10^{-3}$, one can deduce the ratio: 
${\Gamma_{PI}/\Gamma_{SP}}=0.38$.

In the second step, we can use the measured ratio of $\dpl$ to $\dz$ lifetimes to relate the W exchange 
contribution to a standard spectator rate, according to
\begin{equation}
{\tau_\dpl\over\tau_\dz}={\Gamma_{SP}+\Gamma_{WX}+\Gamma_{SL}\over
\Gamma_{PI}+\Gamma_{SL}},
\label{eqDpltoDz}
\end{equation}
where $\Gamma_{SL}$ represents the rate due to semileptonic charm decay 
and $\Gamma_{WX}$ represents any additional contributions 
having to do with W exchange diagrams (including interference between
spectator and W exchange amplitudes).
Using the previous result for $\Gamma_{PI}/\Gamma_{SP}$ and a $\dz$ semileptonic branching 
fraction of 13.4\% (muonic and electronic combined), one can extract
${\Gamma_{WX}/\Gamma_{SP}}=0.26$.

Finally, to estimate the effects of the weak annihilation diagram, one can use the newest data to compare the 
$\ds$ lifetime (where weak annihilation is not Cabibbo suppressed) to the $\dz$ lifetime \cite{dstodz}:
\begin{equation}
{\tau_\ds\over\tau_\dz}=1.05\times{\Gamma_{SP}+\Gamma_{WX}+\Gamma_{SL}\over
\Gamma_{SP}+\Gamma_{WA}+\Gamma_{SL}},
\label{eqDstoDz}
\end{equation}
from which one can derive
${\Gamma_{WA}/\Gamma_{SP}}=0.07$.

Needless to say, these results are only meant to be illustrative of the technique for isolating the various 
decay contributions, and cannot be taken too seriously by themselves. In practice, one must be 
attentive of the many uncertainties that feed into the calculations. In some cases, the results are quite 
sensitive to the input parameters. For example, a variation of $\pm10\%$ in the semileptonic 
branching fraction alone (consistent with measured errors) leads to a range of answers: $\Gamma_{WX}/\Gamma_{SP}$=0.22 to 0.31 and $\Gamma_{WA}/\Gamma_{SP}$=0.04 to 0.11. 
In general, the lesson one should take away is that all of these contributions can be quite 
significant and that the current data on charm lifetimes should provide the basis for better 
understanding in charm decays in the near future.

\subsection{bottom lifetimes}

\begin{figure}[htb] 
\begin{center}
\epsfig{file=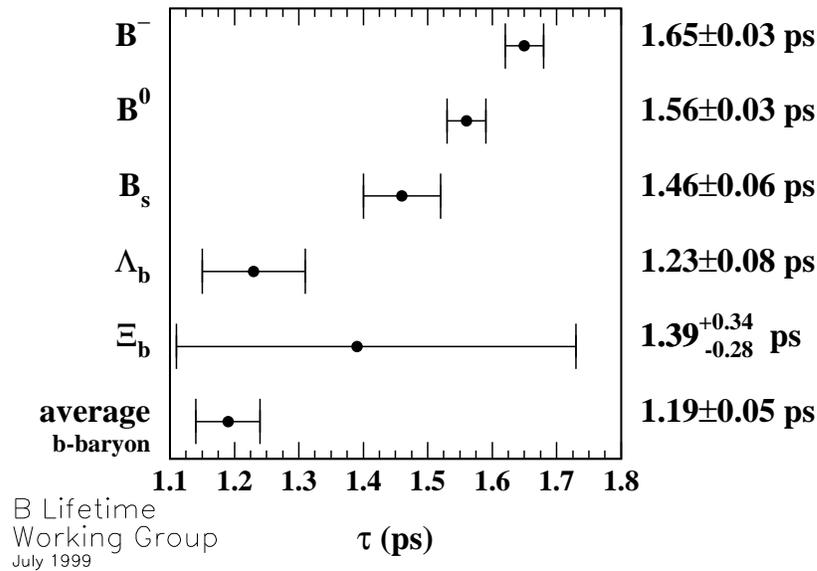,height=3.0in}
\caption[ ]{A summary of lifetimes for weakly decaying bottom particles.}
\label{fgBlifes}
\end{center}
\end{figure}

The current data on bottom decays are not far behind the measurements of charm decays. 
Figure~\ref{fgBlifes} shows the 
most recent results on bottom lifetimes as reported by the B Lifetime Working Group \cite{BlifeWG}. Of 
note is the recent SLD measurement \cite{Abe:1999yj} of $\tau_\bpl/\tau_\bz$, which is the most precise to date. New 
measurements of $\bpl$ and $\bz$ lifetimes are also available from ALEPH \cite{Calderini:1999} and OPAL \cite{Abbiendi:1998av}. Not shown in 
the diagram is the CDF measurement \cite{Abe:1998wi} of the $\bc$ lifetime $\tau_\bc=0.46^{+0.18}_{-0.16}\pm0.03$ ps.


\begin{figure}[htb]
\begin{center} 
\epsfig{file=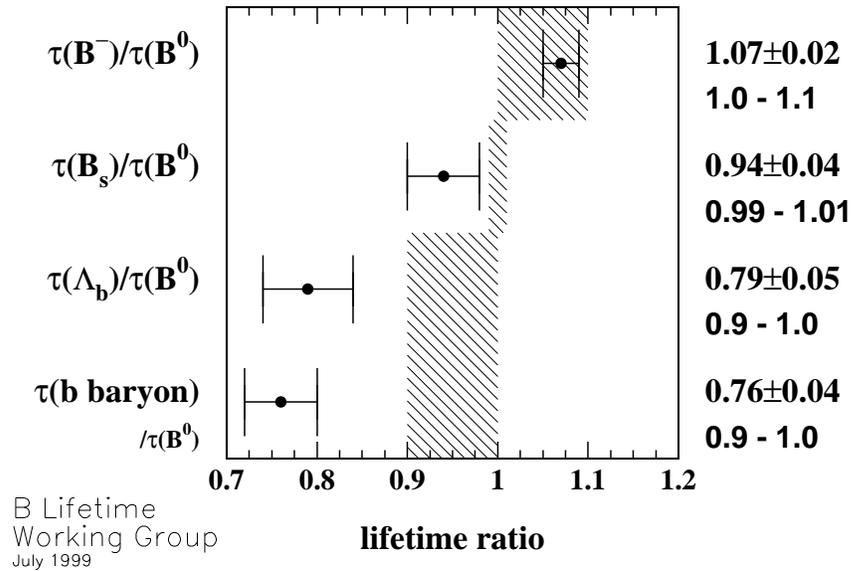,height=3.0in}
\caption[ ]{Measured bottom lifetime ratios compared to predictions (shaded) from reference \protect\cite{Bigi:1994ey}.}
\label{fgBratios}
\end{center}
\end{figure}

The status of the predictions for bottom particle lifetimes is in somewhat better shape than for the charm 
system. Estimates are usually based on the operator product expansion \cite{Bellini:1996ra}:
\begin{equation}
\Gamma\approx{G_F^2m_Q^5\over129\pi^3}(A_1+{A_2\over m_Q^2}+
{A_3\over m_Q^3}+O({1\over m_Q^4})),
\label{eqOPE}
\end{equation}
which calculates corrections in powers of one over the heavy quark mass. The $A1$ term 
represents the spectator processes, 
$A2$ parameterizes some differences between the baryons and mesons, and the $A3$ term includes W exchange, 
weak annihilation and Pauli interference effects. In charm decays, the lighter charm quark mass makes 
questionable the use of this expansion, but in 
B decays the correction terms are about 10\% of what they are 
in the charm sector, and this provides a plausible framework for calculation. 
Figure~\ref{fgBratios} shows a comparison 
of experimental measurements and theoretical predictions for several ratios of bottom lifetimes. The shaded 
area shows the predictions of reference \cite{Bigi:1994ey}. For the mesons, there is good agreement between theory and 
experiment, and the measured ratios are close to unity, as expected if the decays are dominated 
by spectator 
contributions. For the baryons, the observer might be inclined to wonder at the discrepancy between theory 
and experiment. It should be noted however that there are other 
predictions for bottom lifetimes \cite{Neubert:1997we} that 
are more conservative and include the range of current measurement. This remains a point of controversy.


The next few years will see important new measurements in this sector. Some of the most interesting 
should come from Run II at the Tevatron. These will provide more precise measurements of the $\bs$ and 
baryon lifetimes that are needed to reach a general understanding of bottom decays.

\section{Meson Mixing}

\begin{figure}[htb] 
\begin{center} 
\epsfig{file=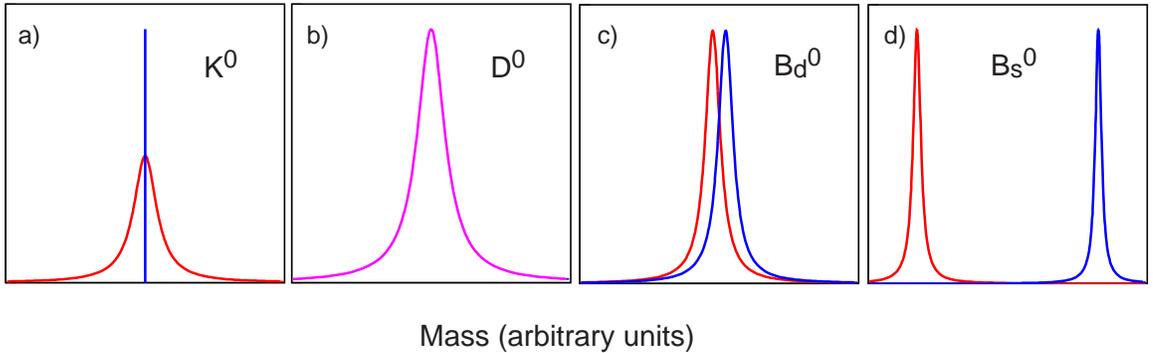,width=6.0in}
\caption[ ]{A schematic overview of neutral meson systems subject to flavor mixing.}
\label{fgneutralmesons}
\end{center} 
\end{figure}

Before launching into the latest results on neutral meson mixing it is useful to begin with a review of the 
properties of the different systems. Figure~\ref{fgneutralmesons} shows schematic mass plots for all four neutral 
systems that are subject to weak flavor mixing \cite{Golowich}. In each case, the scale is artfully chosen to emphasize 
the mass and width differences between the physical eigenstates. Figure~\ref{fgneutralmesons}a shows the neutral kaon 
system, with the broad peak of the $\ks$ and, at essentially the same mass, the narrow spike of the $\kl$, which 
has a decay width 580 times smaller. Figure~\ref{fgneutralmesons}b shows the charm $\dz$ system. Although only 
one curve appears visible, both states of the neutral D have been plotted. The Standard Model predicts an 
immeasurably small difference in width and mass for the two charm D mesons. Figure~\ref{fgneutralmesons}c depicts the $\bd$ 
mesons, with essentially the same widths and a small but noticeable difference in mass. Finally, Figure~\ref{fgneutralmesons}d provides an educated guess for the $\bs$ system. There is expected to be a small difference in width 
between the two mesons (20\% difference in the plot) and a substantial difference  in mass. Interestingly, 
these four systems appear to cover the range of possibilities for mixing.


\begin{figure}[htb] 
\begin{center} 
\epsfig{file=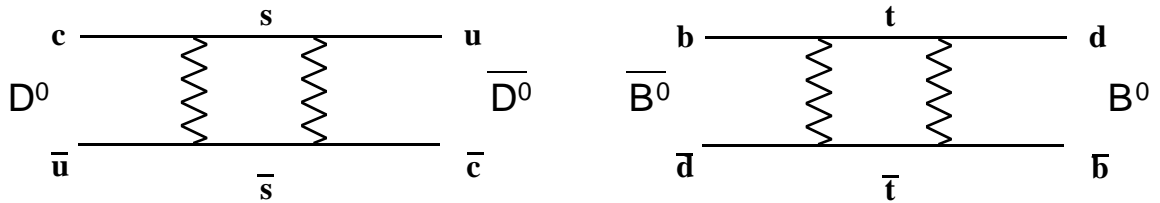,width=6.0in}
\caption[ ]{Representative box diagrams that mediate mixing of charm mesons (left) and
bottom mesons (right).}
\label{fgboxdiags}
\end{center} 
\end{figure}

In order to identify the motivation for studies in meson mixing, it is necessary to understand the source of 
some of the differences in Figure~\ref{fgneutralmesons}. Often, the degree of mixing of a neutral meson system is parameterised 
by
\begin{equation}
\rmix\equiv{\Gamma(M\ra\overline{M}\ra f)\over\Gamma(\overline{M}\ra f)},
\label{eqrmix}
\end{equation}
which describes the rate for a particle to mix and then decay to a particular final state, relative to the rate for 
the particle to decay to that final state without mixing. In order to calculate the probability of mixing, one 
usually begins by calculating the amplitudes of flavor-changing box diagrams such as the ones shown in 
Figure~\ref{fgboxdiags}. On the left is a mixing diagram for the $\dz$ system. Many such box diagrams contribute, 
with different intermediate quark propagators. In the D system, the intermediate propagators are all down-type quarks. In the B system the intermediate propagators are all up-type quarks. 
Calculation for these diagrams shows that the amplitude is proportional to the mass-squared of the 
intermediate quark. Mixing in the B system is therefore dominated by diagrams with heavy internal top 
quarks, and consequently the mixing rate is large in this system ($\rmix\approx 1$). In D mixing, one would 
expect that diagrams with the heavy bottom quarks would dominate, but these are strongly suppressed by 
CKM couplings, and it is actually diagrams with internal strange quarks that dominate \cite{GIM}. Since the 
strange quark mass is so much smaller than the top quark mass, this contribution to D mixing is 
correspondingly smaller than the top quark contribution to B mixing. Moreover, in the charm system it is 
necessary to feed the heavy charm quark 4-momentum through the light strange quark internal propagators, pulling 
them off shell in the process and contributing another suppression factor of the form $m_s^2/m_c^2$. When all 
is said and done, mixing from SM box contributions to the D system are expected to be extremely small, 
leading to $\rmix\approx 10^{-10}$. Other processes may contribute from on-shell intermediate states \cite{onshellDmixing}, nearby resonances \cite{Golowich:1998pz} or 
from penguin diagrams \cite{Petrov:1997ch}, but these too are predicted to be very small. Standard Model mixing in the D 
system is therefore expected to be immeasurable by any current experiment.


The profound difference  between mixing in the bottom system and mixing in the charm system is what 
drives the experimental approach to these systems. In the B system, where mixing from the Standard Model 
is large, mixing measurements are used to study CKM couplings (most notably $\Vtd$). In the D system, 
where SM contributions are small, searches for mixing are used to explore possible contributions from new 
physics.

\subsection{D mixing}

\begin{figure}[ht] 
\begin{center} 
\epsfig{file=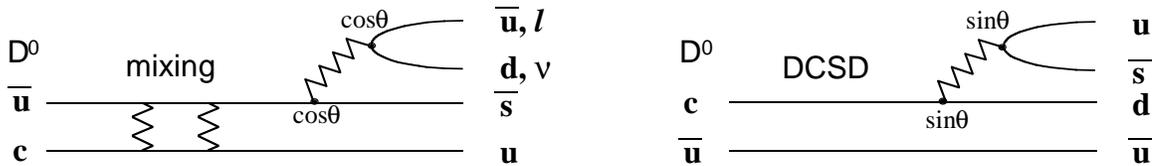,width=6.0in}
\caption[ ]{Two diagrams for ``wrong-sign" decays of D mesons.}
\label{fgwsdiags}
\end{center} 
\end{figure}

The traditional method of observing mixing involves identifying the flavor of the meson both at production 
and at decay. In this way, it is possible to determine if the meson has mixed during the interim. In the 
charm system, the most popular means of tagging the produced D is to reconstruct $\dstarpl$($\dstarmi$) decays to 
$\pipl$$\dz$($\pimi\dzbar$). In this case, the charge of the pion tells whether the initial D meson is a $\dz$ or a $\dzbar$. 
The decay mode of the D can subsequently be used to determine the flavor of the D at decay. As an example, 
the left diagram in Figure~\ref{fgwsdiags} shows how a $\dz$ can mix via a box diagram into a $\dzbar$, which then decays via a spectator 
process to $\kpl\pimi$ or $\kpl l^-\overline\nu$. Experimentally, if the sign of the reconstructed kaon is the same as the sign 
of the pion from the $\dstar$ decay then the event is termed a ``wrong-sign" event, and is a candidate for D 
mixing. 


Unfortunately, for hadronic final states, there are two means for producing wrong-sign events. The first 
involves mixing, as in the left plot in Figure~\ref{fgwsdiags}. The second is doubly-Cabibbo-suppressed decay, as in the right plot of Figure~\ref{fgwsdiags}. Although the 
DCS rate is expected to be only about 1\% of the Cabibbo-favored decay rate, it is an enormous background 
when compared to the extremely small mixing signal expected. Therefore, the wrong-sign rate for hadronic 
final states is a combination of three terms: mixing, DCS decay, and interference between the two. Equation 
\ref{eqhadmixtime} shows the time evolution of hadronic wrong-sign decays in the 
limit of small mixing \cite{smallmix}:
\begin{equation}
\Gamma(\dz\ra\kpl\pimi)\propto e^{-\Gamma t}[4|\lambda|^2+
(\Delta M^2+{\Delta\Gamma^2\over 4})t^2+
(2Re\lambda\Delta\Gamma+4Im\lambda\Delta M)t],
\label{eqhadmixtime}
\end{equation}
where $\lambda$ quantifies the relative strength of DCS and CF amplitudes.
The first term, proportional to $e^{-\Gamma t}$, represents the pure DCS decay rate. The second term, 
proportional to $t^2e^{-\Gamma t}$ represents mixing, which can have contributions from both mass and width 
differences of the eigenstates. The third term, proportional 
to $te^{-\Gamma t}$, represents the interference between 
mixing and DCS amplitudes. In contrast, wrong-sign semileptonic final states are not produced by DCS 
decays, and the time evolution of those states is simply:
\begin{equation}
\Gamma(\dz\ra\kpl l^-\nu)\propto e^{-\Gamma t}
(\Delta M^2+{\Delta\Gamma^2\over 4})t^2
\label{eqsemimixtime}
\end{equation}
This mode is obviously cleaner theoretically, but more challenging experimentally because of the missing 
neutrino.

To set the scale for the latest measurements it is useful to review some previous results. One of the most recent (and least ambiguous) limits on mixing comes from a study at the FNAL 
E791 experiment \cite{Aitala:1996vz}, which examines semileptonic final states. In that study, the 90\% C.L. limit on 
mixing is $\rmix<0.50\%$, a value that is typical of current measurements. There is also an older CLEO 
measurement \cite{Cinabro:1993nh} that gives a wrong-sign signal of 
$\rws=0.77\pm0.25\pm0.25\%$. However, since that result 
did not discriminate between DCS and mixing (there was no vertex chamber for measuring decay lengths in 
the old detector), it is popularly attributed to DCS decays.

\begin{figure}[htb] 
\begin{center} 
\epsfig{file=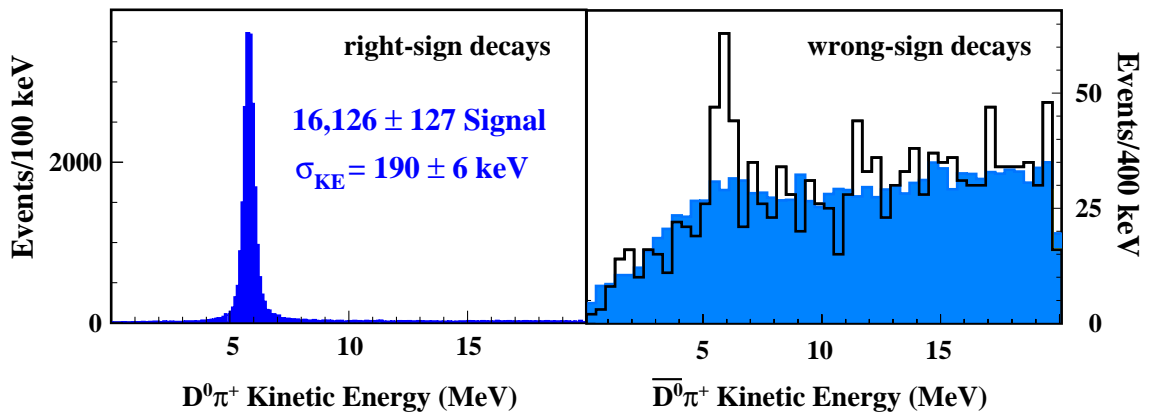,width=6.0in}
\caption[ ]{CLEO kinetic energy distribution of right-sign $\dstarpl\ra\dz\pipl$ and c.c. decays (left) and wrong-sign $\dstarpl\ra\dzbar\pipl$ and c.c. decays (right). In the wrong-sign plot, the shaded area shows the expected 
background level.}
\label{fgcleomix}
\end{center} 
\end{figure}

This summer, a new CLEO study \cite{Artuso:1999hy} that examines $\dstar\ra\dz\pi\ra({\rm K}\pi)\pi$ decays shows a dramatic 
improvement in sensitivity over the older results. This improvement is driven primarily by two effects: 
excellent mass resolution, which reduces non-$\dstar$ backgrounds, 
and high efficiency at short decay times, 
which helps to distinguish between DCS decays and mixing. 
Figure~\ref{fgcleomix} shows plots of the kinetic energy 
of the $\dz\pi$ system, which should peak at 6 MeV for decays 
from the $\dstar$. About 16000 right-sign signal 
events are apparent, with a mass resolution of 190 keV. 
This impressive resolution is due in part to 
a new trick being used by CLEO analysts. The slow pion from the $\dstar$ decay is required to come from the 
beam ribbon \cite{beamribbon}, providing an extra vertex constraint that is an effective aid to improving the momentum 
resolution.


The right side of the figure shows the CLEO D$\pi$ kinetic energy distribution for the wrong-sign decays, with about 60 events 
in the signal peak. A little more than half of the background comes from $\dz$ to K$\pi$  decays combined with a 
random pion to give a wrong-sign $\dstar$ candidate decay. Smaller background contributions come from other 
charm decays and uds events. From these results, CLEO calculates a wrong-sign ratio of $r_{ws}=0.34\pm0.07\pm0.06\%$.

To disentangle DCS decays from the mixing contribution, the decay time distribution is fit to the three terms given in Equation~\ref{eqhadmixtime}.
The results are expressed in terms of the parameters $x'$ and $y'$, 
which are related 
to the mass and width differences ($\deltaM$ and $\deltaGamma$) of the physical eigenstates, 
and the relative phase between DCS and CF decay amplitudes ($\delta$):
\begin{eqnarray}
x' & = & {\deltaM\over\Gamma}\cos\delta+{\deltaGamma\over 2\Gamma}\sin\delta \nonumber \\
y' & = & {\deltaGamma\over 2\Gamma}\cos\delta-{\deltaM\over\Gamma}\sin\delta.
\label{eqxprimeyprime}
\end{eqnarray}
The CLEO 95\% C.L. limits are reported as $|x'|<3.2\%$ and $-5.9\%<y'<0.3\%$.
Assuming that the phase angle $\delta$ is approximately zero \cite{phase}, one can 
relate these limits to 95\% C.L. limits on $\rmix$ due to 
non-zero $\deltaGamma$($\rmix < 0.17\%$ \cite{ylimit}) and non-zero $\deltaM$ ($\rmix < 0.05\%$).

The reader may note that the $\rmix$ limit from the constraint on $x'$ 
is an order of magnitude more restrictive than 
previous measurements. The limit from $y'$ 
is not as restrictive only because the central value for $y'$ is 
about 1.8 standard deviations 
away from zero. Although this discrepancy is not terribly significant, it is 
interesting in its own right.
Recently, a direct search by E791 \cite{deltaGamma}
for non-zero $\deltaGamma$ has also been
performed by looking for a difference in lifetimes for
decays to different final states, yielding a sensitivity to $\deltaGamma$
comparable to the new CLEO limit.
Future work along the same lines can be expected from several experiments.

This subject will be pursued vigorously in the next few years. The FOCUS experiment at FNAL has 
already shown \cite{Sheldon:1999kc} preliminary results on the decay $\dstar$ 
to $\dz$ to K$\mu\nu$. 
From this mode alone, FOCUS expects to be able to set a limit on $\rmix$ of about $\rmix<0.12\%$ if there is 
no indication of mixing. In the near future, B factories will also contribute significantly to these studies. A 
design luminosity year at BaBar will produce about $10^7$ $\dstar\ra\dz\pi$ decays, which should also lead to 
some interesting results.

\subsection{B mixing}

\begin{figure}[htb]
\begin{center} 
\epsfig{file=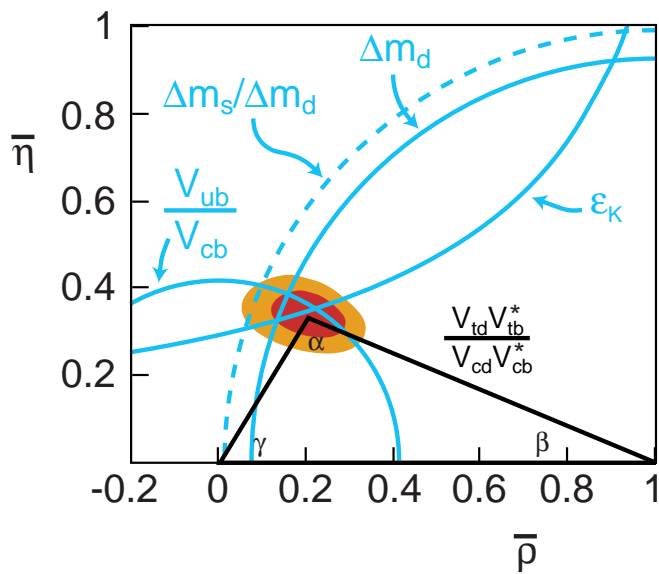,height=3.0in}
\caption[ ]{
A summary of constraints on the unitarity triangle. Measured central values of 
$\epsilon_{\rm K}$, $\Vub/\Vcb$, and $\deltaMd$ are shown as solid lines without
their uncertainties. 
The 95\% C.L. lower limit on $\deltaMs$ is shown as the dashed line.
Likelihood contours for the apex of the triangle come from reference \protect\cite{Parodi:1999nr}.
}
\label{fgutriangle}
\end{center} 
\end{figure}

As was suggested earlier, the implied purpose to studying B mixing is to explore CKM parameters, $\Vtd$ in 
particular. This is easily illustrated with 
Figure~\ref{fgutriangle}, which shows the triangle corresponding to the CKM unitarity condition 
$\Vud\Vubstar+\Vcd\Vcbstar+\Vtd\Vtbstar=0$. The apex of the triangle is constrained 
 by measurements of 
$\Vub/\Vcb$ from semileptonic B decays, measurements of CP violation in the kaon system, measurements 
of $\deltaMd$ from B mixing, and
the lower limit on $\deltaMs$ from $\bs$ mixing. The length of the upper right side of the triangle is 
given by $\Vtd\Vtbstar/\Vcd\Vcbstar$. 
Since $\Vtb$ is close to unity, $\Vcd$ is well-measured from charm decays, 
and $\Vcb$ is approximately $-\sin\theta_c$ 
in the SM; $\Vtd$ remains the limiting factor in determining the length of that side of the triangle. A precise 
measure of that side of the triangle would provide excellent complementary information to the angle 
measurements expected from B factory measurements. Checking consistency in the set of 
measurements that over-constrain the triangle is of great interest in the search for new physics. In particular, 
new measurements of $\Vtd$ and $\sin2\beta$ could constrain the apex of the triangle independently of other data.


Equation~\ref{eqdelMd} \cite{Altarelli:1988zf} shows the relationship between the mass difference $\deltaMd$ measured from $\bd$ mixing 
and the CKM parameter $\Vtd$. Although the measurement of $\deltaMd$ is now quite good (a total of 26 
measurements have been made from LEP, SLD and CDF \cite{BOSC}), the theoretical uncertainties on the 
many coefficients in Equation~\ref{eqdelMd} lead to roughly a 20\% uncertainty in $\Vtd$. 
However, simultaneous measurements of $\bd$ and $\bs$ mixing can give much better precision on $\Vtd$ 
through the ratio of mass differences shown in Equation~\ref{eqdelMdMs} \cite{Buras:1997th}. In this case, many uncertainties cancel 
and there remains about a 5\% theoretical uncertainty on the extraction of $\Vtd$. 

\begin{equation}
\deltaMd={G_F^2\over6\pi^2}m_\bd f_\bd^2B_\bd\eta_{QCD}F(m_t^2)|\Vtd\Vtbstar|^2
\label{eqdelMd}
\end{equation}

\begin{equation}
{\deltaMs\over\deltaMd}={m_\bs f_\bs^2B_\bs\over m_\bd f_\bd^2B_\bd}
\left|{\Vts\over\Vtd}\right|^2=
(1.15\pm0.05)^2\left|{\Vts\over \Vtd}\right|^2
\label{eqdelMdMs}
\end{equation}

To date, $\deltaMs$ has not been measured. Evidence of $\bs$ mixing is clear, but only lower limits on $\deltaMs$ have been determined. Nonetheless, constraints on the unitarity triangle from 
other measurements suggest that it may be just beyond the current measured limits. Assuming the SM, the 
contours in Figure~\ref{fgutriangle} show the present estimate of the apex of the unitarity triangle. The central measured 
value of $\deltaMd$ suggests the apex should lie on the solid quarter circle centered 
at $(\overline\rho,\overline\eta)=(1,0)$. 
The limit on $\deltaMs$ ($\deltaMs>12.4~\psecinv$ is used in the figure) corresponds to the dashed quarter circle just 
outside the $\deltaMd$ curve. Higher limits on $\deltaMs$ push the circle to smaller radius, further constraining the apex of the triangle.

Once again, the identification of mixed events involves tagging the flavor of the meson both at production 
and at decay, and measuring the time evolution of mixing. 
The mixing frequency determines $\Delta M$.
Several techniques have been utilized for $\bs$ decays. The initial state can be tagged by 
examining the charge of leptons or kaons in the opposite hemisphere, by examining an associated kaon in 
the same hemisphere, by calculating a weighted jet charge for either jet, or by using the jet angles in the 
case of polarized beams. The decaying meson can be tagged by using charged leptons in 
the final state, by using partially or fully reconstructed D mesons, or by reconstructing two vertices in the 
decay hemisphere (associated with bottom and charm decay). 

\begin{figure}[htb] 
\begin{center} 
\epsfig{file=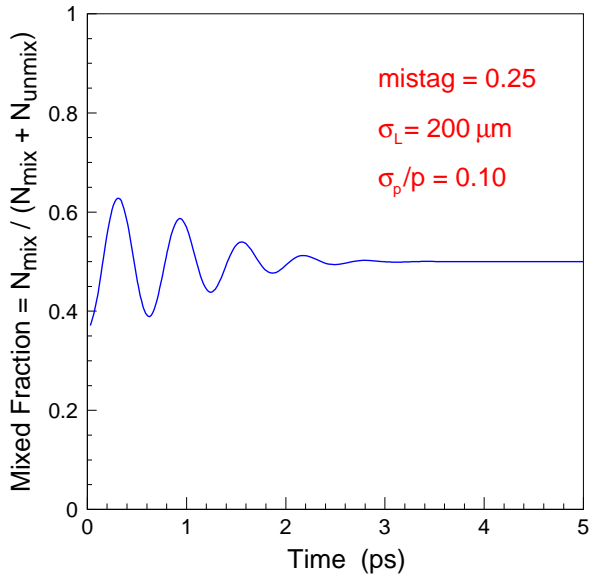,height=3.0in}
\caption[ ]{An idealized measurement for $\bs$ mixing using an infinite data sample of pure $\bs$ decays, with $\deltaMs=10~\psecinv$.}
\label{fgidealBs}
\end{center} 
\end{figure}

The four keys to a precise measure of $\deltaMs$ are excellent proper time resolution, high purity of the $\bs$ 
decay sample (the worst backgrounds tend to come from other B decays), accurate
tagging of the initial and final state mesons, and as always, high statistics. In 
order to illustrate the challenge to experimentalists, Figure~\ref{fgidealBs} shows an idealized experiment with infinite 
statistics and no background for $\deltaMs=10~\psecinv$. 
Even in this case, the measurement is not easy. The vertical axis measures the 
fraction of events identified as mixed, as a function of the proper decay time of the $\bs$. In a perfect 
experiment, this curve should start at zero and oscillate between zero and one. The oscillation never 
makes it all the way to zero or all the way to one because the mistag rate (25\% in the figure) 
dilutes the measurement. 
This effect is exacerbated by smearing due to decay length resolution
(200 $\mu$m in the figure). 
At higher values of proper decay 
time, the amplitude of the oscillation degrades because of increased uncertainty on the decay time due to the 
boost resolution (10\% assumed in the figure).


\begin{figure}[htb] 
\begin{center} 
\epsfig{file=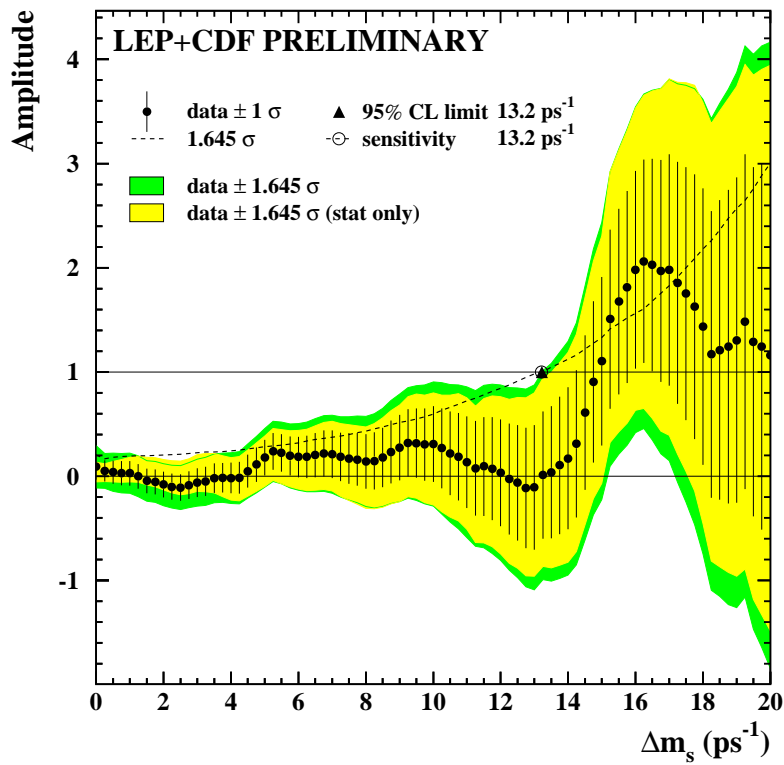,height=4.0in}
\caption[ ]{Fitted amplitude for mixing versus the $\bs$ mass difference. LEP and CDF data only.}
\label{fgmslepcdf}
\end{center} 
\end{figure}

Figure~\ref{fgmslepcdf} shows the combined results on $\deltaMs$ from LEP and CDF. In order to 
understand this plot, it is necessary to recognize that the probability for mixing is proportional to 
$1-\cos\deltaMs t$, where $t$ is the $\bs$ decay time. The figure shows the results of fits for many different values of 
$\deltaMs$ to oscillation data from many experiments. For each point, the data are fit to a function 
proportional to $1-A\cos\deltaMs t$, where the oscillation amplitude $A$ is a fit parameter. If mixing occurs at that particular value 
of $\deltaMs$, then the fitted value of $A$ should be unity. At other values of $\deltaMs$, the fitted value of A should 
be close to zero, consistent with no oscillation. In short, the plot can be thought 
of as a Fourier analysis of oscillation data, with the vertical axis showing the amplitude $A$ as a function of 
frequency. The limit on $\deltaMs$ from this plot alone is $\deltaMs>13.2~\psecinv$ at 95\% C.L.


\begin{figure}[htb] 
\begin{center} 
\epsfig{file=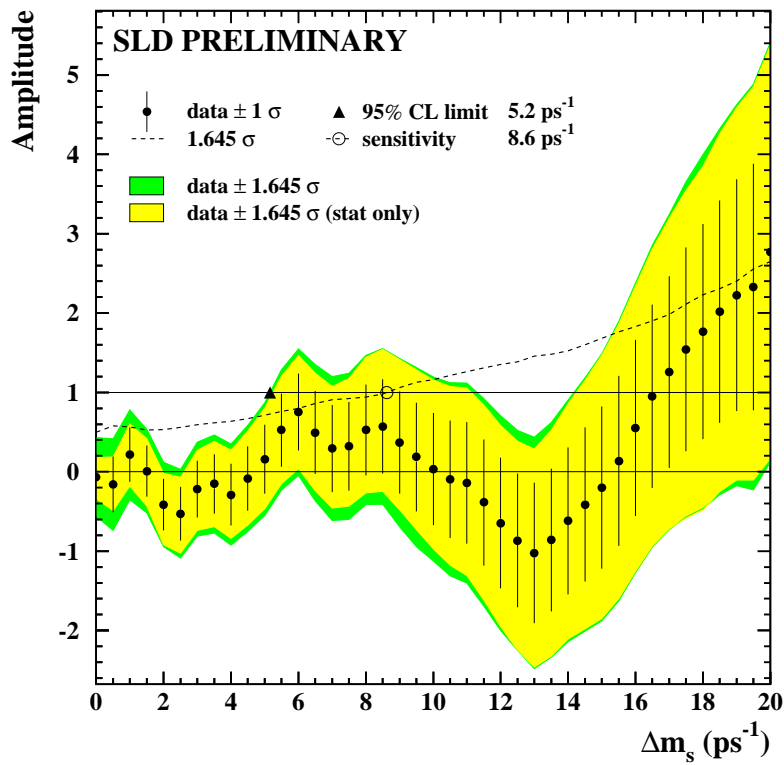,height=4.0in}
\caption[ ]{ Fitted amplitude for mixing versus the $\bs$ mass difference. SLD results only.}
\label{fgmssld}
\end{center} 
\end{figure}

Figure~\ref{fgmssld} is an analogous plot for new data from SLD \cite{Abe:1999yk}. 
These results show dramatic improvement since 
Moriond 99, driven primarily by new tracking and substantial improvements in decay length resolution. 
Three separate analyses are employed, corresponding to reconstructed final states of a charm vertex plus 
lepton, a high-momentum lepton, or a pair of vertices displaced from the primary vertex. Two analyses not 
shown, but expected in the summer of 2000, search for $\bs$ decays using final states that include an 
exclusively reconstructed $\ds$ decay, or a lepton and charged kaon. Once again, the plot shows the fitted 
value of the oscillation amplitude $A$ as a function of $\deltaMs$. At low values of $\deltaMs$ the uncertainty in 
$A$ is about a factor of two larger than the results from combined CDF and LEP data. However, at high values 
of $\deltaMs$, the uncertainties are comparable, so that these data make a significant contribution to the world 
average.


\begin{figure}[htb] 
\begin{center} 
\epsfig{file=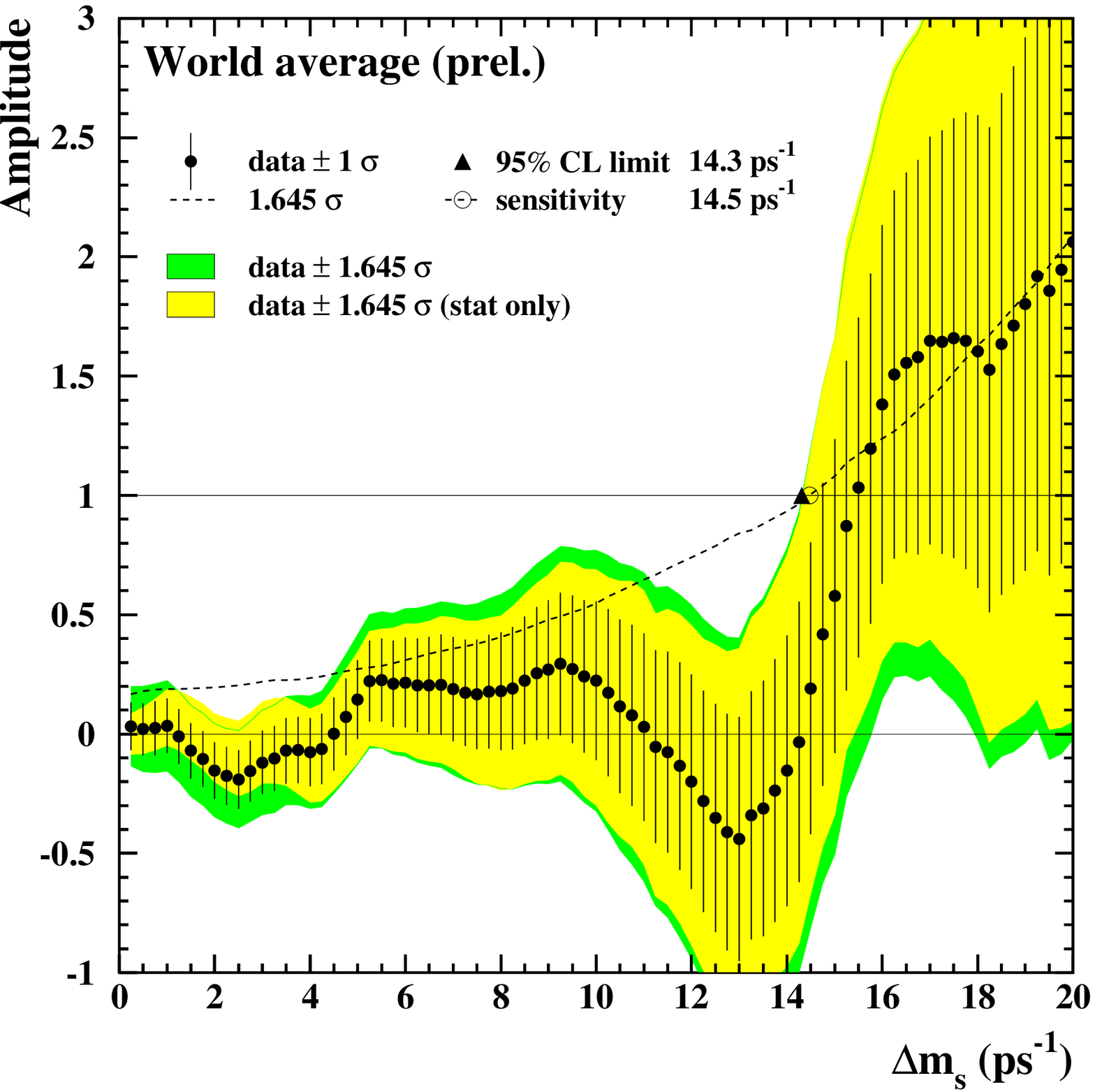,height=4.0in}
\caption[ ]{ Fitted amplitude for mixing versus the $\bs$ mass difference. Combined world data.}
\label{fgmsall}
\end{center} 
\end{figure}

Figure~\ref{fgmsall} shows the combined data from all experiments, updated as of December 99 \cite{BOSC}. A total of 11 analyses contribute. The 95\% C.L. 
lower limit on $\deltaMs$ from these data is $\deltaMs>14.3~psecinv$, up from 12.4~$\psecinv$ reported at EPS 99 (Tampere). 
The new higher limit on $\deltaMs$ provides further constraint on the unitarity triangle of Figure~\ref{fgutriangle}. 
The dashed circle from the $\deltaMs$ limit now moves just inside the curve that represents the central value of 
$\deltaMd$. This change clips off a significant fraction of the previously allowed area for the apex of the 
triangle. 


\begin{figure}[htb] 
\begin{center}
\epsfig{file=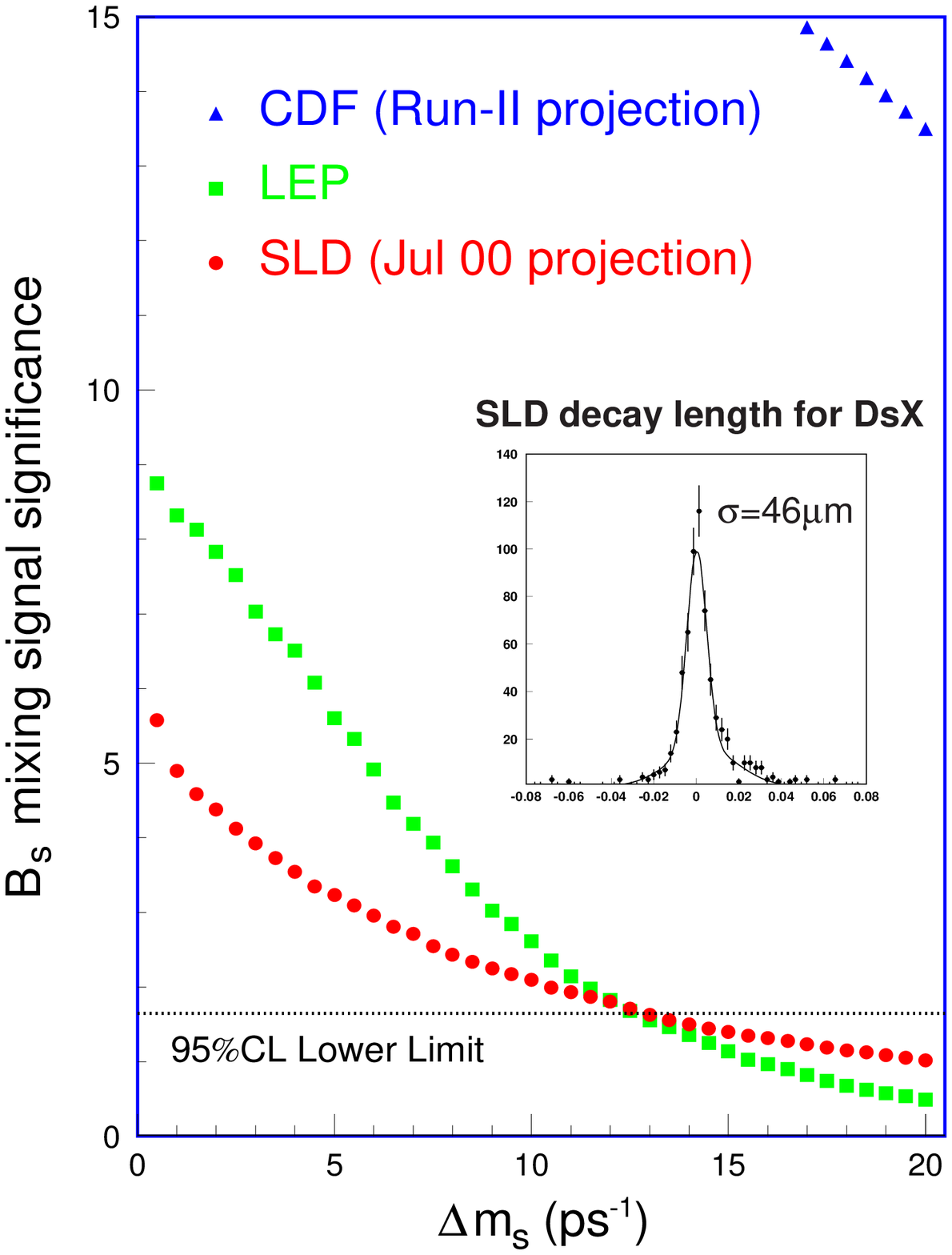,height=4.0in}
\caption[ ]{Sensitivity to $\bs$ mixing as a function of the $\bs$ mass difference for three different experiments. The CDF curve (triangles) crosses
the 95\% C.L. line around $\deltaMs=50~\psecinv$.}
\label{fgBsfuture}
\end{center}
\end{figure}

In the next few years, further work in this area should prove very interesting. Figure~\ref{fgBsfuture} gives an idea of 
what we should expect from future studies of $\bs$ mixing. The vertical scale of the plot is called ``significance" and 
is the inverse uncertainty in the amplitude parameter $A$ of Figures \ref{fgmslepcdf}-\ref{fgmsall}. It can therefore be interpreted as  
the analyzing power for discriminating between $A=0$ and $A=1$. The squares map the significance 
as a function of $\deltaMs$ from the combination of all the LEP experiments. The circles 
plot the significance from the SLD data set when all five analyses are included. Both curves cross the 95\% C.L. limit at about $\deltaMs$=13 ps$^{-1}$.


The comparison of these two curves is particularly interesting in two respects. First of all, it is surprising 
that they appear on the same graph when one considers that the LEP data 
sample represents 40 times more luminosity than the SLD data sample. Secondly, the shapes of the two 
curves are significantly different. The SLD curve is dramatically flatter than the LEP curve, and at high 
values of $\deltaMs$ the SLD significance even wins out over LEP. The reason for both these features 
is the very precise vertex resolution of the SLD detector. This allows SLD researchers to do more inclusive 
style analyses, that are more efficient, in order to compete with the statistics from LEP. It also allows 
SLD to retain good sensitivity to the very fast oscillation at high values of $\deltaMs$. The insert in the figure 
shows the vertex resolution achieved by the ongoing analysis that tags $\bs$ decays via a fully reconstructed 
$\ds$ decay. The 46 $\mu$m resolution for the central Gaussian (60\% of the area) is roughly four times better than the resolution achieved 
in a typical LEP study.

A competitive race will continue between LEP and SLD for the next year or so as each group tries to improve 
the sensitivity of the measurements. This will be done in the hopes of actually seeing the $\bs$ 
oscillation, which is predicted by the other measurements of Figure~\ref{fgutriangle} as being just beyond the limits of 
current analysis. However, if the oscillation is not seen at LEP or SLD in the next year, new players will soon
dominate the field. The triangles in the upper right corner of Figure~\ref{fgBsfuture} show 
what to expect from the CDF experiment after Run II. That curve crosses the 95\% C.L. line around 
$\deltaMs$=50 ps$^{-1}$. Assuming the Tevatron experiments can trigger efficiently on displaced vertices, those 
experiments will dominate the $\bs$ oscillation measurements in just a few years. It is especially interesting to note that the future Tevatron data
should either confirm the SM estimate of $\Vtd$ 
or prove that the SM is incorrect 
because the triangle 
doesn't close (see also Michael Peskin's comment at the end of this paper).

\section{CP Violation}

The recent turn-on of two new B factories has focussed a lot of attention on
the already hot topic of CP violation. This year there are two items relevant
to CPV in the B system that are worthy of note. Both of these topics are covered
by other speakers at this conference, so the summary here will be brief.
The first is an update on searches for
CPV in $\bz\ra\psi\ks$ (see also M. Paulini's contribution to these proceedings).
The second is a pair of measurements of $\Gamma(\bz\ra\pipl\pimi)$ 
and $\Gamma(\bz\ra\kpl\pimi)$ from CLEO that
has implications for future B factory measurements of the mixing angle $\alpha$
(see also R. Poling's contribution to these proceedings).

It is common knowledge that CP violation in a decay rate is due to the interference between two or more 
amplitudes with different CP-conserving and different CP-violating phases. In particular, if both 
amplitudes are pure decay amplitudes, then the CP violation is called ``direct CPV". In this case, CP 
violation is constant in time and can be measured via integrated asymmetries. On the other hand, if one 
of the amplitudes involves mixing then the violation is called ``indirect CPV", and the asymmetry evolves 
in time. In the charm system, where mixing is expected to be negligible, the search for CP violation is 
generally a search for direct CP violation in integrated asymmetries. In the bottom system, where mixing is 
large, time dependent asymmetries are used to quantify indirect CP violation. 

In the bottom system, measurements of CP asymmetries are used primarily to explore the CKM matrix. 
From a theorist's view point, final states that are dominated by tree level amplitudes and are CP 
eigenstates are the cleanest modes for extracting the CKM parameters. Two popular examples of such 
modes are $\bz\ra\psi\ks$ (which is often considered the golden mode for measuring $\beta$) 
and $\bz\ra\pipl\pimi$ (which is often talked about as a method to measure $\alpha$). Both of these modes have played 
important roles in recent results.

\begin{figure}[htb] 
\begin{center}
\epsfig{file=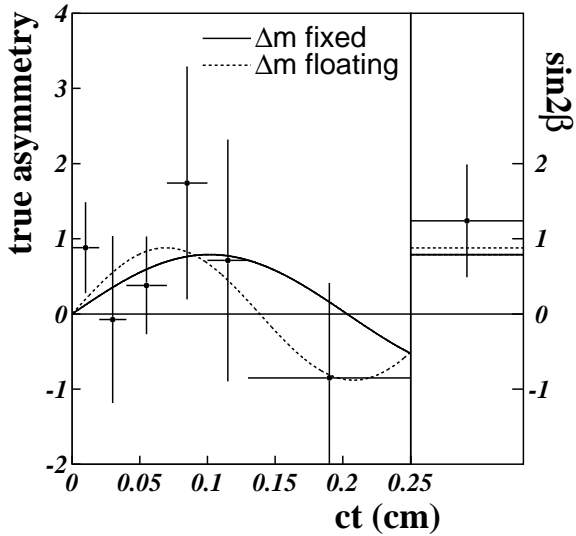,height=3.0in}
\caption[ ]{CDF fits to time-dependent (left) and integrated (right) CP asymmetries in $\bz\ra\psi\ks$ decays.}
\label{fgpsiKs}
\end{center}
\end{figure}

This year, the CDF experiment has updated a CP asymmetry measurement \cite{Affolder:1999gg} of $\bz$($\bzbar$) decays to $\psi\ks$ and a new result from 
ALEPH \cite{ALEPHpsiKs} for the same final state has also become available.
Preliminary indications from CDF were available already last year, but the signal was marginal, and this year 
researchers have worked hard to squeeze out the last bit of sensitivity from the data. Of the roughly 400
reconstructed $\psi\ks$ events in the current CDF data sample, about half of them occur within the acceptance of 
the vertex detector, where the decay lengths are well-measured. These events are used to measure a time 
dependent asymmetry in search of CP violation. The other half of the events do not have well-measured 
decay times and are used in the measure of a time integrated asymmetry. Figure~\ref{fgpsiKs} shows the time 
dependence of the $\bz$/$\bzbar$ asymmetry with the best fit for an oscillation on the left, and the single data 
point of the time integrated asymmetry on the right. The two results are combined to get a measure of 
$\sin2\beta_{CDF} = 0.79^{+0.41}_{-0.44}$. ALEPH performs a similar time-dependent asymmetry measurement on
23 well-reconstructed decays to get $\sin2\beta_{ALEPH} = 0.93^{+0.64+0.36}_{-0.88-0.24}$. Together, the two experiments constrain
$\sin2\beta$ to be greater than zero with 98.5\% probability. 
Although this result does 
not provide a very meaningful test of the previous constraints on the unitarity triangle, it does provide an 
indication of CPV and some reassurance that this measurement will be a good target for B factory studies in 
the near future.


The other interesting results this year of relevance to CPV are new measurements \cite{Kwon:1999hx} from the CLEO 
collaboration for the branching fractions of $\bz\ra\pipl\pimi$ and 
$\bz,\bpl\ra$K$\pi$. 
Since the K$\pi$ final states are believed to be dominated by penguin diagrams, 
these decays offer a measure of the importance of penguin contributions.
CLEO measurements for B$\ra$K$\pi$ range from 1.2 to 1.9$\times 10^{-5}$,
slightly larger than B$\ra\pi\pi$ branching fractions, which are 
believed to be dominated by
tree diagrams. The relatively large
K$\pi$  branching fractions therefore indicate that penguin diagrams play an 
important role in these decays. In particular, one can use the 
measured $\bz\ra\kpl\pimi$  and $\bz\ra\pipl\pimi$ rates to get a 
rough estimate of the $\pipl\pimi$ penguin amplitude relative to the $\pipl\pimi$ tree amplitude. Following the method 
outlined in the BaBar physics book \cite{Harrison:1998yr}, section 6.1.2, one comes to:
\begin{equation}
0.25<{A_{penguin}^{\pi\pi}\over A_{tree}^{\pi\pi}}<0.57.
\label{eqpenguintotree}
\end{equation}

Although the precise numerical result should not be taken too seriously, it does point out that penguin 
amplitudes are likely to be significant in the $\pipl\pimi$ decay mode.
Consequently, the study of CPV in the $\pipl$$\pimi$ final state must include interference between tree 
amplitudes, mixing amplitudes, and penguin amplitudes. This naturally makes the extraction of $\alpha$ much 
more difficult. As has been pointed out by London and Gronau \cite{Gronau:1990}, the $\pipl$$\pimi$ asymmetry can still be used 
in combination with branching fraction measurements of $\bpl\ra\pipl\piz$ and $\bz\ra\piz\piz$ to measure $\alpha$, 
but this is a considerably harder problem, with new ambiguities. The general conclusion is that 
measurements of $\alpha$ at the B factories will be a challenge.

\section{Summary}

This paper has examined four topics of recent research in heavy quark decays.
In each case, interesting new results are available this year, and these point the way to even better results in the
near future.
First of all, new measurements of the $\ds$ lifetime provide useful data for 
improving our understanding in the mechanisms of charm decay. 
In the near future, precision measurements of charm baryon
lifetimes from FNAL fixed target experiments FOCUS and SELEX should help
complete that understanding. Secondly, results from a new CLEO search for charm mixing
are just released, which improve the sensitivity to charm mixing by about an
order of magnitude. In the next few years, efforts at FOCUS and at the B factories will further the search. Thirdly, attempts to measure the $\bs$ mixing
frequency have improved this year, resulting in a higher limit on $\deltaMs$. Efforts at
LEP and SLD will continue for at least another year, with the hope of 
seeing the oscillation. If it is not found in the next year, Run II data
from the Tevatron experiments is expected to extend the reach in
$\deltaMs$ by more than a factor of two. In time, this will either confirm the 
estimates of $\Vtd$, or it will point to an
interesting conflict within the Standard Model. Finally, efforts have begun
to measure CP asymmetries in the B system. New results at CDF and ALEPH suggest
that $\sin2\beta$ is within the expected range and should be an easy target
for B factory measurements. On the other hand, 
measurements of $\alpha$ via CP asymmetries of $\bz\ra\pipl\pimi$
may prove to be more difficult in light of the new 
CLEO measurements of
the $\bz\ra$K$\pi$ and $\bz\ra\pi\pi$ branching fractions, which suggest that penguin contributions play an important role in these decay
modes.



\def\Discussion{
\setlength{\parskip}{0.3cm}\setlength{\parindent}{0.0cm}
     \bigskip\bigskip      {\Large {\bf Discussion}} \bigskip}
\def\speaker#1{{\bf #1:}\ }

\Discussion

\speaker{Michael Peskin (SLAC)}
There is a comment that is implicit in your discussion of 
Figure~\ref{fgutriangle}, but it is 
nice to make it explicit.
There is one leg of the unitarity triangle which is determined by $V_{ub}$.
This determination is independent of possible new physics.
The long leg on the right is determined by $\Delta m_s/\Delta m_d$.
If indeed  $\Delta m_s$ turns out to be of the order of 
16 ps$^{-1}$, then these two measurements
would already give an accurate 
determination of the unitarity triangle in the context 
of the standard CKM model.
On the other hand, this suggestion may turn out to be wrong.
But if you made the right
hand leg about half that  size, it would not be possible to make a triangle.
 That is, if   $\Delta m_s$ turns out to be greater than 
about 35 inverse picoseconds, the CKM model is wrong or at least incomplete.
You noted that the CDF sensitivity goes far beyond this value, up to about
 50  ps$^{-1}$.  So when CDF comes into the game, we might determine the
CKM triangle within the Standard Model, or we might be able to rule out the 
Standard 
Model just on the basis of the  $\Delta m_s$  information.

\speaker{Jon Thaler (University of Illinois)}
Ron Poling presented a measurement of $\gamma$
 that is about two $\sigma$ away from your favored value.
 Do you have any comments on this discrepancy?

\speaker{Blaylock}
Since I first encountered this question at the conference, I have had time to
better understand the assumptions underlying the
CLEO analysis \cite{Kwon:1999hx} that Ron presented. That estimate of $\gamma$ depends upon a fit to 14 charmless decay modes of the B mesons. By assuming factorization 
of amplitudes, these decay modes are fit to five parameters. For decay 
modes that are dominated by spectator diagrams, an argument might be
made in favor of factorization since the two ends of the W connect to two
separated quark currents (though there is some controversy even about this point). 
However, most of the modes used in the fit have large penguin contributions, 
making a factorization argument unreasonable. In my mind it is not surprising that
the CLEO fit does not yield the same result for the weak phase $\gamma$ as other
estimates.


\begin{thebibliography}{99}

\bibitem{suppression}
In helicity suppression, spin zero meson decay to a relativistic, back-to-back quark-antiquark pair is suppressed by angular momentum conservation. In color suppression, the final state quarks are 
required to carry the correct color charge so that the net final state is colorless. For the diagrams of Figures 
\ref{fglifediags}c to \ref{fglifediags}f, 
this is an important constraint. 
For the diagrams of \ref{fglifediags}a and \ref{fglifediags}b, 
correct color matching is already 
guaranteed by the couplings to the colorless W boson.

\bibitem{Caso:1998}
C.~Caso {\it et al.},
Eur.\ Phys.\ J.\  {\bf C3}, 1 (1998).

\bibitem{Aitala:1998ca}
E.M.~Aitala {\it et al.}
[Fermilab E791 Collaboration],
Phys.\ Lett.\ {\bf B445}, 449 (1999)
hep-ex/9811016.

\bibitem{Bonvicini:1999ub}
G.~Bonvicini {\it et al.}
[CLEO Collaboration],
Phys.\ Rev.\ Lett.\ {\bf 82}, 4586 (1999)
hep-ex/9902011.


\bibitem{Cheung:1999tz}
H.W.~Cheung,
hep-ex/9912021.

\bibitem{lifetechnique}
The technique presented here for using charm branching 
fractions to extract $\Gamma_{PI}$, $\Gamma_{WX}$ 
and $\Gamma_{WA}$ was first suggested to me by H. Cheung, reference \cite{Cheung:1999tz}.

\bibitem{dstodz}
The factor of 1.05 in Equation~\ref{eqDstoDz} accounts for several small
differences in the decay of $\ds$ and $\dz$ mesons. Purely leptonic decays
of $\ds\ra l\nu$ lower the $\ds$ lifetime by a few percent, and the presence of 
Pauli Interference in Cabibbo suppressed $\ds$ decays raises the
$\ds$ lifetime by a few percent, as do SU(3) breaking effects.
For a complete discussion, see \cite{Bigi:1994ey}.

\bibitem{BlifeWG}
http://home.cern.ch/$\sim$claires/lepblife.html

\bibitem{Abe:1999yj}
K.~Abe {\it et al.}
[SLD Collaboration],
hep-ex/9907051.

\bibitem{Calderini:1999}
G.~Calderini  [ALEPH Collaboration],
CERN-OPEN-99-266.

\bibitem{Abbiendi:1998av}
G.~Abbiendi {\it et al.}
[OPAL Collaboration],
hep-ex/9901017.

\bibitem{Abe:1998wi}
F.~Abe {\it et al.}
[CDF Collaboration],
Phys.\ Rev.\ Lett.\ {\bf 81}, 2432 (1998)
hep-ex/9805034
;
F.~Abe {\it et al.}
[CDF Collaboration],
Phys.\ Rev.\ {\bf D58}, 112004 (1998)
hep-ex/9804014.

\bibitem{Bellini:1996ra}
G.~Bellini, I.~Bigi and P.J.~Dornan,
Phys.\ Rept.\ {\bf 289}, 1 (1997).

\bibitem{Bigi:1994ey}
I.I.~Bigi and N.G.~Uraltsev,
Z.\ Phys.\ {\bf C62}, 623 (1994)
hep-ph/9311243.

\bibitem{Neubert:1997we}
M.~Neubert and C.T.~Sachrajda,
Nucl.\ Phys.\ {\bf B483}, 339 (1997)
hep-ph/9603202.

\bibitem{Golowich}
This way of looking at the various flavor mixing systems was 
first shown to me by Gene Golowich, 
private communication.

\bibitem{GIM}
Incidentally, if one keeps all the strange and down quark box diagram contributions to D mixing, the net 
amplitude is proportional to $(m_s^2-m_d^2)/M_W^2$. If $m_s$ and $m_d$ were equal, this amplitude would vanish 
entirely. This is an example of the GIM mechanism at work.

\bibitem{onshellDmixing}
E.~Golowich,
hep-ph/9706548
;
F.~Buccella, M.~Lusignoli and A.~Pugliese,
Phys.\ Lett.\ {\bf B379}, 249 (1996)
hep-ph/9601343
;
J.F.~Donoghue, E.~Golowich, B.R.~Holstein and J.~Trampetic,
Phys.\ Rev.\ {\bf D33}, 179 (1986).

\bibitem{Golowich:1998pz}
E.~Golowich and A.A.~Petrov,
Phys.\ Lett.\ {\bf B427}, 172 (1998)
hep-ph/9802291.

\bibitem{Petrov:1997ch}
A.A.~Petrov,
Phys.\ Rev.\ {\bf D56}, 1685 (1997)
hep-ph/9703335.

\bibitem{smallmix}
If the reader is only familiar with mixing in the B system, this formula may be
a surprise. 
In the limit of small 
mixing, the familiar term $e^{-\Gamma t}\sin^2\Delta mt$ 
from B mixing is reduced to $e^{-\Gamma t}\Delta m^2t^2$. In the 
charm system, new physics may contribute to a width difference as strongly as it does to a mass difference, so the $\Delta\Gamma^2t^2$ term appears as well.

\bibitem{Aitala:1996vz}
E.M.~Aitala {\it et al.}
[Fermilab E791 Collaboration],
Phys.\ Rev.\ Lett.\ {\bf 77}, 2384 (1996)
hep-ex/9606016.

\bibitem{Cinabro:1993nh}
D.~Cinabro {\it et al.}
[CLEO Collaboration],
Phys.\ Rev.\ Lett.\ {\bf 72}, 1406 (1994).

\bibitem{Artuso:1999hy}
M.~Artuso {\it et al.}
[CLEO Collaboration],
hep-ex/9908040.

\bibitem{beamribbon}
The beam ribbon constraint is only valid for prompt charm events that originate at the primary vertex. For 
that purpose, the CLEO analysis requires a minimum D momentum in order to insure that the D did not 
come from a B decay.

\bibitem{phase}
In recent years, fits to charm decay data 
using certain models have suggested that the 
phase angle between CF and 
DCS decay amplitudes is small 
(for an overview see 
T.~E.~Browder and S.~Pakvasa,
Phys.\ Lett.\  {\bf B383}, 475 (1996)
[hep-ph/9508362]),
but this claim remains controversial. 

\bibitem{ylimit}
The limits due to non-zero $\deltaGamma$ were calculated 
from the lower limit of the CLEO constraint on $y'$,
$-5.9\%<y'$, with $\rmix=y'^2/2$. 

\bibitem{deltaGamma}
E.M.~Aitala {\it et al.}
[Fermilab E791 Collaboration],
Phys.\ Rev.\ Lett.\ {\bf 83}, 32 (1999)
hep-ex/9903012.

\bibitem{Sheldon:1999kc}
P.D.~Sheldon,
hep-ex/9912016.

\bibitem{Parodi:1999nr}
F.~Parodi, P.~Roudeau and A.~Stocchi,
hep-ex/9903063.

\bibitem{Altarelli:1988zf}
G.~Altarelli and P.J.~Franzini,
Z.\ Phys.\ {\bf C37}, 271 (1988).

\bibitem{BOSC}
To view the latest summary of measurements of $\deltaMd$, and a compilation of searches
for $\bs$ mixing, see the B Oscillation
Working Group page at 
http://www.cern.ch/LEPBOSC/.

\bibitem{Buras:1997th}
A.J.~Buras,
hep-ph/9711217.


\bibitem{Abe:1999yk}
K.~Abe {\it et al.}
[SLD Collaboration],
{\it Contributed to 19th International Symposium on Lepton and Photon
                  Interactions at High-Energies (LP 99), Stanford, CA, 9-14
                  Aug 1999}.

\bibitem{Affolder:1999gg}
T.~Affolder {\it et al.}
[CDF Collaboration],
hep-ex/9909003.

\bibitem{ALEPHpsiKs}
Details on the ALEPH analysis of $\bz\ra\psi\ks$ can be found in the note 
ALEPH 99-099, available at 
http://alephwww.cern.ch/ALPUB/oldconf/oldconf\_99.html.

\bibitem{Kwon:1999hx}
Y.~Kwon {\it et al.}
[CLEO Collaboration],
hep-ex/9908039.

\bibitem{Harrison:1998yr}
P.F.~Harrison and H.R.~Quinn
[BABAR Collaboration],
{\it Papers from Workshop on Physics at an Asymmetric B Factory (BaBar
                  Collaboration Meeting), Rome, Italy, 11-14 Nov 1996,
                  Princeton, NJ, 17-20 Mar 1997, Orsay, France, 16-19 Jun
                  1997 and Pasadena, CA, 22-24 Sep 1997}.

\bibitem{Gronau:1990}
M.~Gronau and D.~London,
Phys.\ Rev.\ Lett.\  {\bf 65}, 3381 (1990).

\end{thebibliography}
\end{document}